\documentclass[twocolumn,showpacs,amsmath,amssymb,aps,prl,nobalancelastpage]{revtex4}

\usepackage{amsthm}
\usepackage{newlfont}
\usepackage{graphicx}
\usepackage{subfigure}
\usepackage[section]{placeins}
\usepackage{psfrag}
\usepackage{caption2}
\usepackage{flafter}
\usepackage{bm}%
\def\fnum@figure{\figurename\thefigure}
\renewcommand{\figurename}{Fig.}

\begin{document}


\title{A generating mechanism for higher order rogue waves}
\author{\mbox{}\hspace{-1cm}J.S.He $^{1}$, H.R. Zhang $^{1}$, L.H. Wang $^{1}$,
K. Porsezian $^{2}$}\author{A.S.Fokas$^{3,4}$ }
\address{$^{1}$Department of Mathematics, Ningbo University, Ningbo, Zhejiang 315211 P.R. China.
\\ \mbox{}\hspace{-1.5cm}$^{2}$Department of Physics, Pondicherry University, Puducherry  605014 India
\\ \mbox{}\hspace{-.5cm}$^{3}$School of Engineering and Applied Sciences, Harvard
University, Cambridge, MA 02138,USA.
\\ \mbox{}\hspace{-.3cm}$^{4}$Permanent address:DAMTP, University of Cambridge, Cambridge, CB3 0WA, UK}
\begin{abstract}
We introduce a mechanism for generating  higher order rogue waves
(HRWs) of the nonlinear Schr\"odinger(NLS) equation: the progressive
fusion and fission of $n$ degenerate breathers associated with a
critical eigenvalue $\lambda_0$ creates an order-$n$ HRW. By
adjusting the relative phase of the breathers in the interacting
area, it is possible to obtain different types of HRWs. The value
$\lambda_0$ is a zero point of an eigenfunction of the Lax pair of
the NLS equation and it corresponds to the limit of the period of
the breather tending to infinity. By employing  this mechanism we
prove two conjectures regarding the total number of peaks, as well
as a decomposition rule in the circular pattern of an order-$n$ HRW.
\\
 \noindent KEYWORDS:  rogue wave, breather, nonlinear Schr\"odinger
 equation.
\end{abstract}
\pacs{ 42.65.Tg,42.65.Sf,05.45.Yv,02.30.Ik \mbox{}\vspace{-0.5cm}}
 \maketitle
\noindent{\bf Introduction.} Rogue waves(RWs) in the ocean are
catastrophic natural phenomena with a long history and  fascinating
mariner stories \cite{list}. Detailed studies of RWs have occurred
only during the past five decades \cite{draper,dean,Kharif,osborne}.
A prototype one dimensional rogue wave is the so-called Peregrine
soliton \cite{peregrine}; this soliton exhibits the two remarkable
characteristics of first-order RWs: (a) localized behavior in both
space and time, (b) the existence of one dominant peak. RWs have
been observed in several fields, including  optics
\cite{soli1,soli2,dudley}, superfluid helium \cite{ganshin},
Bose-Einstein condensates \cite{konotop1}, plasmas \cite{ruderman,
shukla}, microwaves \cite{hohmann}, capillary phenomena\cite{shats},
telecommunication data streams \cite{vergeles}, and inhomogeneous
media \cite{arecchi}.

A typical modeling equation for RWs in fiber optics is the
celebrated nonlinear Schr\"odinger (NLS) equation \cite{nlseq},
\begin{equation}\label{nlse}
  i q_t+q_{xx}+2|q|^2q=0.
\end{equation}
Here $q=q(x,t)$ is a complex smooth function of $x$ and $t$, and the
subscripts denote partial derivatives. The Peregrine soliton
\cite{peregrine}, which is the first-order RW \cite{akhmediev1} of
the NLS equation, has been observed experimentally in fibers
\cite{akhmediev2}, in a water tank \cite{akhmediev3}, and in
multi-component plasmas \cite{bailung}. Recently, a super rogue wave
\cite{akhmediev4}, i.e., a second order RW, has also been observed
in a water tank. In addition to the NLS equation, the Hirota
equation \cite{akhmediev5,he1}, the first-type derivative NLS(DNLS)
equation \cite{he2}, the third type DNLS equation \cite{he3}, the
NLS-Maxwell-Bloch equations \cite{he4}, the discrete NLS equation
\cite{akhmediev6a}, the two-component NLS equations
\cite{guo1,fabio1,zhangweiguo}, and the Davey-Stewartson equation
\cite{kalla,jianke}, also admit RWs. These results show that RWs may
be generic phenomena in nonlinear systems.

 The Peregrine soliton \cite{peregrine,akhmediev1} of the NLS equation is
expressed in terms of a simple rational formula; it corresponds to a
simple profile and can be obtained from a breather solution via the
simple limit of the period of modulation approaching infinity.
However, higher order rogue waves(HRWs) \cite{gaillard1,gaillard2,
gaillard3,gaillard4} are expressed in terms of complicated formulas
and their profiles exhibit
 several different  interesting patterns \cite{akhmediev6,guo2,akhmediev7,
 jianke2,akhmediev8}. These patterns include a fundamental pattern
consisting of a simple central highest peak surrounded by several
gradually decreasing peaks [see Figs. 2(a) and 3 in
Ref.\cite{akhmediev8}], an equal-height triangular pattern [see Fig.
2(b) in Ref. \cite{akhmediev8} and Fig.2 in  Ref. \cite{guo2}], and
circular pattern [see Fig.4 in  Ref. \cite{akhmediev8}].

Taking into consideration the complexity of the relevant formulas
\cite{akhmediev9, he5}, as well as the plethora of different
possible patterns, it is a challenging problem to elucidate the
mechanism of HRW generation. There exist two important conjectures
regarding HRWs.
\begin{itemize}
\item In the case of a single fundamental pattern, an order-$n$ RW
has $n(n+1)-1$ non-uniform peaks \cite{akhmediev8}; in the case when
there exist several patterns, an order-$n$ RW has  $n(n+1)/2$
uniform peaks \cite{gaillard2,gaillard3}.
\item In the case when an order-$n$ RW displays a ring structure,
 the outer ring has $2n-1$ uniform peaks, and the inner structure
is an order-($n-2$) RW\cite{akhmediev8}.
\end{itemize}

In this work, we present a generating mechanism for HRWs of the
NLS equation, and using this mechanism, we prove the above two
conjectures. Furthermore, we discuss several new interesting
patterns of HRWs.

\noindent{\bf A Degenerate n-fold DT  and inverse DT.} In order to
study the breather and the RW solutions of the NLS equation, we
shall use the determinant representation of the Darboux
transformation (DT) introduced in \cite{neugebauer, matveev,he6}.
Furthermore, we shall use the notations and  the main results of
these references regarding the $n$-fold DT (theorem 1 in \cite{he6})
and the related functions ($q^{[n]},r^{[n]},\phi^{[n]}$) generated
by the n-fold DT(corollary 1 in \cite{he6}). In order to satisfy the
reduction requirement $q^{[n]}=-(r^{[n]})^*$,we choose
$f_{2k}=(-f^*_{2k-1\ 2}, f^*_{2k-1\ 1})^T, k=1, 2, \cdots, n$, where
 $T$ denotes matrix transposition and the asterisk denotes complex
conjugation. Under this reduction, $q^{[n]}$ is a solution of the
NLS equation generated by an $n$-fold DT starting with the seed
solution $q$. In the following we always use this reduction
condition.

Theorem 1 and corollary 1 cited above imply that an $n$-fold DT
$T_n$ of the NLS equation annihilates its independent generating
functions, which are the eigenfunctions $f_i(i=1,3,5\cdots,2n-1)$
associated with $n$ distinct eigenvalues
$\lambda_1,\lambda_3,\lambda_5, \cdots,\lambda_{2n-1}$. This means
that if we fix the given set of eigenvalues, we cannot apply DTs
more than once. Recall that the formulas for the eigenfunctions
$f_i(i=1,3,5\cdots,2n-1)$ differ only by the fact that they involve
different eigenvalues $\lambda_i$. However, in order to obtain a HRW
for a critical eigenvalue $\lambda_0$, we must apply repeated DTs.
This difficulty can be overcome by noting that the annihilated
eigenfunctions can be re-created by taking  the
 limit $\lambda_i \rightarrow \lambda_1$ of the
eigenvalues used in the DT \cite{guo2}. We set $f_i=\phi(\lambda_i)$
and $f^{[n]}_i=\phi^{[n]} |_{\lambda=\lambda_i}$. We shall use the
determinant representation of the $n$-fold DT [see Eq. (14) of
\cite{he6}] to illustrate the relevant construction. It is
straightforward to verify that  $f_1^{[1]}=0$,
 and hence  we cannot apply the DT again with the eigenvalue
$\lambda_1$. Let $\lambda_3=\lambda_1+\epsilon$; then
\begin{align*}
&f_3^{[1]}\mbox{\hspace{-0.8cm}}&=&f^{[1]}_3(\lambda_1+\epsilon)=f^{[1]}_1(\lambda_1+\epsilon) \\
&\mbox{\hspace{-0.8cm}} &=&f^{[1]}_1(\lambda_1)+(
 \dfrac{\partial f^{[1]}_1(\lambda_1+\epsilon)}{\partial
 \epsilon}|_{\epsilon=0}\ )\epsilon +O(\epsilon).
\end{align*}
Hence, the limit
$$\lim_{\epsilon \rightarrow 0}\dfrac{1}{\epsilon}
f_3^{[1]} =\dfrac{\partial f^{[1]}_1(\lambda_1+\epsilon)}{\partial
\epsilon}|_{\epsilon=0} \triangleq  f^{[1]}_{1}$$
yields  a
transformed eigenfunction associated with $\lambda_1$, which can be
used to generate a new DT  so that we can apply this DT with the
given eigenvalue $\lambda_1$ for a second time. Similarly, set the
second degenerate eigenvalue  $\lambda_5=\lambda_1 +\epsilon$ in
$f_5^{[2]}$; the limit
$$\lim_{\epsilon \rightarrow
0}\dfrac{1}{\epsilon^2} f_5^{[2]} =\dfrac{\partial^2
f^{[2]}_1(\lambda_1+\epsilon)}{\partial \epsilon^2}|_{\epsilon=0}
\triangleq  f^{[2]}_{1}$$ re-creates a transformed eigenfunction
associated with $\lambda_1$ of the two-fold DT. Note that the zero
order and the first order terms of $\epsilon$ in $f_5^{[2]}$ yield
zero contributions. In general, for an $n$-fold DT, we can use the
following theorem on  $\phi^{[n]}(\lambda)$ and $q^{[n]}$ using  the
degenerate limit $\lambda_i \rightarrow \lambda_1$ by a similar
analysis, based on the determinant representation given by Theorem 1
and Corollary 1 of \cite{he6}. The following notations, including
matrix elements $[(t_1)_{12}]_{ij}$ and $(W_{2n})_{ij}$,
are given in \cite{he6}.\\
{\bf  Theorem 1}{\sl \ An n-fold DT with a given eigenvalue
$\lambda_1$ is realized by the  degenerate limit $\lambda_i
\rightarrow \lambda_1$. This degenerate n-fold DT yields the
transformed eigenfunction $\phi^{[n]}$ of $\lambda$,where
\begin{equation}\label{rwnegf}
    \phi^{[n]}=\frac{1}{|W'_{2n}|}\begin{pmatrix}
                                \begin{vmatrix}
                                  \hat{\phi}(n) & \lambda^n\phi_1 \\
                                  W'_{2n} & \hat{\xi}'_{2n-1} \\
                                \end{vmatrix}\\
                                \noalign{\vskip5pt}
                                \begin{vmatrix}
                                  \hat{\phi}(n) & \lambda^n\phi_2 \\
                                  W'_{2n} & \hat{\xi}'_{2n} \\
                                \end{vmatrix} \\
                              \end{pmatrix},
\end{equation}
as well as   a new solution $q^{[n]}$ of the NLS equation starting
with the seed solution $q$,where
\begin{equation}\label{degenerateddtqn}
    q^{[n]}(x,t;\lambda_1)=q-2i\frac{|Q'_{2n}|}{|W'_{2n}|},
\end{equation}
with
\begin{align}
    &W'_{2n}&=\left(\frac{\partial^{n_i-1}}{\partial\varepsilon^{n_i-1}}\bigg|_{\varepsilon=0}
    (W_{2n})_{ij}
    (\lambda_1+\epsilon)\right)_{2n\times2n},\nonumber \\
    &\hat{\xi}'_{2n-1}&=
       \left(\frac{\partial^{n_i-1}}{\partial\varepsilon^{n_i-1}}
       \bigg|_{\varepsilon=0}\hat{\xi}_{2n-1,i}(\lambda_1+\epsilon)\right)_{2n\times1},\nonumber \\
    &\hat{\xi}'_{2n}&=
       \left(\frac{\partial^{n_i-1}}{\partial\varepsilon^{n_i-1}}
       \bigg|_{\varepsilon=0}\hat{\xi}_{2n,i}(\lambda_1+\epsilon)\right)_{2n\times1},\nonumber \\
    &Q'_{2n}&=\left(\frac{\partial^{n_i-1}}{\partial\varepsilon^{n_i-1}}\bigg|_{\varepsilon=0}
    (Q_{2n})_{ij} (\lambda_1+\epsilon)\right)_{2n\times2n}, \nonumber
\end{align}
$n_i=[\frac{i+1}{2}]$, $[i]$ denotes the floor function of $i$,
$Q_{2n}$ is the determinant in the numerator of $(t_1)_{12}$\cite{he6}}. \\
Starting with different seed solutions $q$, Eq.
(\ref{degenerateddtqn}) yields different degenerate solitons and
 breathers.  Furthermore, by choosing a special
 eigenvalue $\lambda_1=\lambda_0$ associated with $\phi(\lambda_0)=0$ ,
  Eq. (\ref{degenerateddtqn}) yields an order $n$ RW. In the latter case,
all orders of derivatives with respect to $\epsilon$ in $\phi^{[n]}$
and $q^{[n]}(x,t;\lambda_1)$ are increased  by $1$ because
$\phi(\lambda_0)=0$. The main idea of the above procedure for
constructing rogue wave is the following: According to the
determinant representation in Theorem 1 and Corollary 1 of
\cite{he6}, there are two degenerate cases
 in $T_{2k}$, i.e., $\lambda_i \rightarrow \lambda_1$ and
$f_i=\phi(\lambda_i)=0(i=1,3,\cdots,2k-1)$. It is easy to recognize
that $q^{[2k]}$ generated by $T_{2k}$ is given by an indeterminate
form $\dfrac{0}{0}$ in the above degenerate cases. Thus, whether
$\lambda_i=\lambda_1+\epsilon $ or $\lambda_i=\lambda_0+\epsilon $ ,
smooth solutions can be obtained by higher-order Taylor expansion in
determinants with respect to $\epsilon$ as in Theorem 1.

In order to get an order-$(n-2)$ RW from an order-$n$ RW by a simple
limit, it is necessary to use an inverse DT. For a general
eigenvalue $\lambda$, the $x$ part of the Lax pair of the NLS
equation admits the solution $\phi(\lambda)$, as well as the
linearly independent solution $\psi(\lambda)=(\psi_1,\psi_2)^T$.
Furthermore $\psi^{[n]}=T_n \psi$ and $\phi^{[n]}=T_n\phi$ are
linearly independent because $T_n$ is a linear transformation of
$\psi$ and $\phi$. Let $g_k \triangleq (g_{k1},g_{k2})^T
=\psi(\lambda_k)$; then the Wronskian determinant $W(f_i,g_i)=f_{i1}
g_{i2}-f_{i2} g_{i1}$ of $f_i$ and $g_i$ is a non-zero constant.
Using the determinant representation of the one-fold DT,
 $T(\lambda;f_1,f_2)$, generated by $f_1$ and $f_2$, we find the transformed
 functions
\begin{align*}
&g_1^{[1]}= \dfrac{(\lambda_1-\lambda_2)W(f_1,g_1)}{|W_2|}
                              \begin{pmatrix}
                              f_{21} \\
                              f_{22} \\
                              \end{pmatrix},\\
&g_2^{[1]}= \dfrac{(\lambda_1-\lambda_2)W(f_2,g_2)}{|W_2|}
                              \begin{pmatrix}
                              f_{11} \\
                              f_{12} \\
                              \end{pmatrix},
                              \end{align*}
which are not zero in contrast to $f_1^{[1]}=0$ and $f_2^{[2]}=0$.
Hence, we can use $g_1^{[1]}$ and $g_2^{[1]}$ to generate the second
fold DT $T(\lambda;g_1^{[1]},g_2^{[2]})$. Using a straightforward
calculation with the help of Theorem 1 in \cite{he6}, it can be
shown that the two-fold DT is given by
\begin{align}
&\mbox{\hspace{-0.3cm}}T_2=T(\lambda;g^{[1]}_1,g^{[1]}_2)
T(\lambda;f_1,f_2)=(\lambda-\lambda_1)(\lambda-\lambda_2)I,
\end{align}
where $I$ is the unit matrix of size 2. Here we present only the
calculation of the element $(T_2)_{11}$. First note that
\begin{align*}
&|W_{4}(g_1,g_2,f_1,f_2)|=-(\lambda_2-\lambda_1)^2W(f_1,g_1)W(f_2,g_2);\\
&(\tilde{T}_2)_{11}\mbox{\hspace{-0.1cm}}
=\mbox{\hspace{-0.1cm}}-(\lambda-\mbox{\hspace{-0.1cm}}\lambda_1)
(\lambda-\mbox{\hspace{-0.1cm}}\lambda_2)
(\lambda_2-\mbox{\hspace{-0.1cm}}\lambda_1)^2W(f_1,g_1)W(f_2,g_2).
\end{align*}
 Hence,
\begin{equation*}
 (T_2)_{11}\mbox{\hspace{-0.1cm}}=\mbox{\hspace{-0.1cm}}
\dfrac{(\tilde{T}_2)_{11}}
{|W_{4}(g_1,g_2,f_1,f_2)|}=(\lambda-\lambda_1)(\lambda-\lambda_2).
\end{equation*}
Thus $T(\lambda;g^{[1]}_1,g^{[1]}_2)$ is the inverse DT of
$T(\lambda;f_1,f_2)$. In general, for an ($n-2$)-fold DT $T_{n-2}$
generated by {$f_1,f_2,\dots,f_{2n-5},f_{2n-4}$}, we can find a
one-fold inverse DT as  follows(note that $g^{[n-2]}_{2n-3}=T_{n-2}
g_{2n-3}$ or $g^{[n-2]}_{2n-2}=T_{n-2} g_{2n-2}$ and
$f^{[n-2]}_{2n-3}$or $f^{[n-2]}_{2n-2}$ are linearly independent):\\
{\noindent\bf Theorem 2}{\sl\ Let the $(n-1)$-th fold DT be
$T(\lambda;f^{[n-2]}_{2n-3},f^{[n-2]}_{2n-2})$ after an (n-2)-fold
DT $T_{n-2}$, and
$g^{[n-1]}_{2n-3}=T(\lambda;f^{[n-2]}_{2n-3},f^{[n-2]}_{2n-2})g^{[n-2]}_{2n-3},
g^{[n-1]}_{2n-2}=T(\lambda;f^{[n-2]}_{2n-3},f^{[n-2]}_{2n-2})g^{[n-2]}_{2n-2}$.
Then the n-th fold DT $T(\lambda;g^{[n-1]}_{2n-3},g^{[n-1]}_{2n-2})$
is the inverse
of the  $(n-1)$-th fold DT. }\\
In other words, the n-th fold DT
$T(\lambda;g^{[n-1]}_{2n-3},g^{[n-1]}_{2n-2}) $  maps $q^{[n-1]}$ to
$q^{[n-2]}$.  This gives an important
connection between RWs  of order ($n-1$) and order ($n-2$).\\
\noindent{\bf Higher order breathers and rogue waves} The first-
order breather of the NLS equation is a periodic traveling wave.
This solution, via the limit of the period approaching infinity,
gives the first-order RW \cite{peregrine,akhmediev1}. However, it is
still not clear how to generate HRWs from multi-breathers, even for
second-order RWs \cite{akhmediev7}. Moreover, the collision of three
breathers \cite{akhmediev10} does not provide  a satisfactory
explanation for the  appearance of different patterns of order-3
RWs.

On the ($x,t$) plane, because of the conservation of the number of
breathers, there exist $n$ separate peaks in each row before and
after the interaction of $n$ breathers. The interaction area is
localized near the origin of the plane between the two closest rows
(or periods) possessing $n$ separate peaks. When the breathers are
in the interaction area, their peaks get closer. Based on the
detailed investigation of the interaction of breathers, we claim the
following mechanism for the generation
 of HRWs:
the progressive fusion and fission of $n$ degenerate breathers
associated with a critical eigenvalue $\lambda_0$ creates an order-
$n$ HRW. Furthermore,  by adjusting the relative phase of the
breathers in the interacting area, it is possible to obtain
different patterns of HRWs. Here $\lambda_0$ is a zero point of the
eigenfunction $\phi(\lambda)$, i.e., $\phi(\lambda_0)=0$, which
corresponds to  the limit of the period of the breathers becoming
 infinitely large. The
relative phase can be adjusted via the tuning of the parameters
$s_i$ in the eigenfunctions $f_i$.

In this work, we shall take a periodic seed $q=ce^{ \mathbf{i}\rho}$
with $\rho=ax+(2c^2-a^2)t$. The corresponding eigenfunctions
$\phi(\lambda)=(\phi_1,\phi_2)^T$ are  given by
             \begin{equation}\label{e2.3}
                 \phi(\lambda)\mbox{\hspace{-0.1cm}}=\mbox{\hspace{-0.2cm}}\left(\mbox{\hspace{-0.1cm}}\begin{array}{c}
                      c e^{\mathbf{i} (\frac{\rho}{2}+d(\lambda))}\mbox{\hspace{-0.1cm}}
                      +\mathbf{i}(\frac{a}{2}+c_1(\lambda)\mbox{\hspace{-0.1cm}}
                      +\mbox{\hspace{-0.1cm}}\lambda) e^{-\mathbf{i}(-\frac{\rho}{2}+
                      d(\lambda))}\\
                       c e^{-\mathbf{i} (\frac{\rho}{2}+d(\lambda))}\mbox{\hspace{-0.1cm}}
                       +\mathbf{i}(\frac{a}{2}+c_1(\lambda)\mbox{\hspace{-0.1cm}}
                       +\mbox{\hspace{-0.1cm}}\lambda) e^{\mathbf{i}(-\frac{\rho}{2}+
                      d(\lambda))}
                       \end{array}\mbox{\hspace{-0.1cm}}
                       \right)\mbox{\hspace{-0.1cm}}.
                 \end{equation}
Here $
c_1(\lambda)=\sqrt{c^2+(\lambda+a/2)^2},d(\lambda)=c_1(\lambda)(x+(2\lambda-a)
t+s_{0}+\Phi), \Phi=\sum_{k=1}^{n-1}s_k\epsilon^{2k}$\cite{guo2},
$n$ denotes the number of steps of the multi-fold DT,
$\lambda_0=-a/2+ic$ is a zero point of the eigenfunction
$\phi(\lambda)$, $\epsilon$ denotes a small parameter when we
consider the degeneracy of the eigenvalues,i.e.,
$\lambda=\lambda_0+\epsilon$, and $s_i$ are complex constants. The
functions $f_i=\phi(\lambda_i)$ have the same form except for the
occurrence of different values of the eigenvalues which is necessary
to generate HRWs via the process of eigenvalue degeneration
$\lambda_i \mapsto\lambda_1$ (see Theorem 1). In the following
examples we set $a=0$. Also, in order to adjust the relative phase
of the breathers in the interaction area according to Theorem 1 and
Corollary 1 of Ref. \cite{he6}, we set $s_i=0(i\geq 1)$, but $s_0$
has different values in different $f_i$. There exist three types of
relative phases of $n$ breathers in the interaction area:
synchronous, anti-synchronous, and quasi-synchronous.

In the interaction area of $n$ synchronous breathers, there exist
progressively increasing fusion via $n-1$ steps from the $n$ lower
peaks to the central maximum peak, and then progressively decreasing
fission via $n-1$ steps from the central maximum peak to the $n$
lower peaks. Here, each step of fusion annihilates one peak and
hence the height of the peaks increases; similarly, each step of
fission creates one new peak and hence the height of the peaks
decreases. These peaks are arranged  as two triangles with one joint
vertex along their perpendicular bisector. Thus, the total number of
non-uniform peaks in the interaction area is $n(n+1)-1$. It is
interesting to note that the outermost row of the interaction area
has $n$ lower peaks, which are close to each other. Hence, the peaks
are much lower than the ones in the nearest row of the
non-interaction area. This phenomenon provides evidence for the
strong interaction of the breathers. When the eigenvalue used in the
breathers approaches the critical value $\lambda_0$,i.e.
$\lambda_i\mapsto \lambda_0$, the periods of all breathers go to
infinity simultaneously, so that only one profile in the interaction
area survives, and this gives the fundamental pattern of a HRW.
Therefore, this pattern of an order-$n$ HRW, has $n(n+1)-1$
non-uniform peaks. The central profile of the three breathers in
Fig. 1 is very similar with  the fundamental pattern of an order-3
RW [see Fig. 3(a) in Ref. \cite{akhmediev8}] of the NLS equation.
The three breathers are plotted according to Theorem 1 and Corollary
1 of \cite{he6} with $a=0.01$,$c=0.5$, $s_0=0$,and $\lambda_1=-0.2+
0.54 \mathbf{i}$ in $f_1$, $\lambda_3=0.1+ 0.55 \mathbf{i}$ in $f_3$
and $\lambda_5=0.03+ 0.56 \mathbf{i}$ in $f_5$.

The interaction of $n$ anti-synchronous breathers is simpler,
although there also exists the fusion or fission of peaks. In the
interaction area, the peaks are closer to each other. By suitable
adjustment of the relative phases of the breathers, $n$ synchronous
breathers become $n$ anti-synchronous breathers, and the
corresponding peaks in the triangle disappear, so that only peaks in
one triangle survive. Specifically, by suitably changing the
relative phase, we observe the disappearance of the lowest peak in
the outermost row of the interaction area, followed by the
disappearance of the two nearest peaks, and so on. This chain
reaction continues until the coalescence of the two triangles. The
collapse of this triangle is stimulated by the loss of the
nearest-neighbor interactions. Thus, there are $n(n+1)/2$ peaks in
the interaction area, which are allocated on the remaining triangle.
If we set $\lambda_i\mapsto \lambda_0$ simultaneously, then the
profile in the interaction  area of the order $n$ breather yields a
triangular pattern of a HRW. Therefore, there are $n(n+1)/2$
equal-height peaks in the triangular pattern of an order-$n$ HRW. A
triangular structure of the three anti-synchronous breathers is
plotted in Fig. 2 by using Theorem 1 of \cite{he6} with $a=0.01$,
$c=0.5$, and $\lambda_1=0.05+ 0.531 \mathbf{i}$ and $s_0=16 $ in
$f_1$, $\lambda_3=0.55 \mathbf{i} $ and $s_0=-20\mathbf{i}$ in
$f_3$, and $\lambda_5=-0.05+ 0.551 \mathbf{i}$ and $s_0=-16 $ in
$f_5$. By suitably choosing different values of the parameters in
$n$ anti-synchronous breathers, so that the relative positions of
the peaks are changed but the total number of peaks is preserved, we
obtain a ring structure associated with the $n(n+1)/2$ peaks in the
interaction area, which gives rise to a circular pattern of an order
$n$ RW in the above limit possessing $n(n+1)/2$ peaks. Figure 3
confirms the ring structure of three anti-synchronous breathers with
$a=0.01$, $c=0.5$, and $\lambda_1=0.05+ 0.54 \mathbf{i}$ and
$s_0=1+\mathbf{i} $ in $f_1$, $\lambda_3=0.55 \mathbf{i} $  and
$s_0=0$ in $f_3$, and  $\lambda_5=-0.05+ 0.56 \mathbf{i}$ and
$s_0=1+\mathbf{i} $ in $f_5$. There exist many other patterns
appearing in the interaction area of $n$ anti-synchronous breathers,
which give rise to several types of HRWs such as two polygons
 (Fig. 4 for an order-5 RW given by Theorem 1 with $\lambda_0=ic,
a=0,c=1/\sqrt{2},s_0=0,s_1=0,s_2=10^6$, $s_3=10 $ for a pentagon or
$s_3= (1+I)\times5\times10^5$ for a heptagon, and $s_4=0$)
 and a triangle in a circle (Fig. 5 for
an order-5 RW given by Theorem 1 with
$\lambda_0=ic,a=0,c=1/\sqrt{2}, s_0=0,
s_1=15,s_2=0,s_3=0,s_4=10^7$).

By suitably choosing the values of the parameters in $f_i$, we
observe $n$ quasi-synchronous breathers from Theorem 1 and Corollary
1 in \cite{he6}. There exist many different patterns in the
interaction area, which implies many interesting types of  HRWs when
$\lambda_i \mapsto \lambda_0$. For example, one can find the
following interesting decomposition of an order-$n$ HRW: an order
-($n-2$) RW surrounded by $2n-1$ peaks \cite{akhmediev8}. The first
nontrivial example of this decomposition is given by the interaction
of  four quasi synchronous breathers. Unfortunately, we are not able
to plot the profile in the interaction area in this case, due to the
complexity of order-$4$ breathers. However, we find two complete
decompositions of the circular pattern: an order-$5$ RW in Fig. 6
with $\lambda_0=ic,a=0,c=1/\sqrt{2},s_0=s_1=s_2=0,s_3=8\times
10^4,s_4=2\times10^7$,
 and an order-$6$ RW in Fig. 7 with
 $\lambda_0=ic,a=0,c=1/\sqrt{2},s_0=s_1=s_2=0,s_3=3\times 10^4,s_4=0,s_5=10^8$.

The inverse DT provides a technique enabling us to prove the above
interesting decomposition rule of an order-$n$ RW. For simplicity we
set $a=0$ in the seed solution, and then we set $\lambda_0=ic$.
Expanding $\phi(\lambda_0+\epsilon)$ with respect to $\epsilon$, we
find that the coefficient of the first-order term in $\epsilon$ is
given by
$$
f_0=\left(
\begin{matrix}
& e^{\mathbf{i} c^2t}(2 \mathbf{i} cx-4c^2t+2\mathbf{i}
cs_{0}+\mathbf{i}) \\
&-e^{-\mathbf{i}c^2t}(2\mathbf{i}
cx-4c^2t+2\mathbf{i} cs_{0}-\mathbf{i})
\end{matrix}
\right),
$$
which is an
eigenfunction associated with $q$ and $\lambda_0$. There exist
another eigenfunction
$$g_0=\left(
\begin{matrix}
e^{\mathbf{i} c^2t} \\
-e^{-\mathbf{i}c^2t}
\end{matrix}
\right)
$$
of $\lambda_0$. Note that $f_0$ and $g_0$ are two linearly
independent eigenfunctions of $\lambda_0$. According to Theorem 1,
an order-$n$ RW is generated by a degenerate $n$-fold DT with the
critical eigenvalue $\lambda_0$ from the periodic seed $q=ce^{2
\mathbf{i} c^2}$. Let $T_{n-1}$ be an ($n-1$)-fold degenerate DT
with $\lambda_0$, so that $g_0^{[n-1]}$ and $f_0^{[n-1]}$ are given
by eq.(\ref{rwnegf}), and are linearly independent. By a tedious
asymptotic analysis we find that $f_0^{[n-1]}=\tilde{f_0}+s_{n-1}
g_0^{[n-1]}$, where $\tilde{f_0}$ is a smooth bounded function.
According to Theorem 2, the $n$-th fold DT defined by $T(\lambda;
g_0^{[n-1]})$ is the inverse of the $(n-1)$-th DT defined by
$T(\lambda; f_0^{[n-2]})$. Thus, by the limit $s_{n-1}\mapsto
\infty$, the $n$-th fold DT $T(\lambda; f_0^{[n-1]})=T(\lambda;
g_0^{[n-1]})$, gives an inverse transform of the $(n-1)$-th fold DT.
Therefore, under this limit, an order-$n$ RW $q^{[n]}$ is reduced to
an order-$(n-2)$ RW, $q^{[n-2]}$. By taking
 $s_{n-1}$ to be  large (but finite), an order $n$ RW is decomposed
  into an order-$(n-2)$ RW and $2n-1$ peaks located on an outer circle
   such that the total number $n(n+1)/2$ of peaks can be realized
   either in a triangular  pattern or in a circular pattern.
The inner order-$(n-2)$ RW can take any of these forms by choosing
$s_i(i=0,1,\cdots,n-3)$. This decomposition rule of  HRWs was
conjectured by Akhmediev and co-workers \cite{akhmediev8}. Figures 5
and 6 show different patterns of the inner lower order RW decomposed
from an order-$5$ RW. The fundamental pattern of the inner order-$3$
RW reduced from an order-$5$ RW is shown in Fig. 4(c) of
\cite{akhmediev8}.  According to this decomposition rule, Figs. 6
and 7 provide the first two non-trivial examples of a complete
decomposition associated with three levels.

In order to show the applicability of the generating mechanism, we
present new types of decomposition of the seventh order, eighth
order  and ninth order RWs in Figs. 8-13. In these figures,
$\lambda_0=ic,a=0, c=1/\sqrt{2}$, and the other non-zero parameters
are $s_6=10^{10} c_0$ in Fig. 8; $s_6=10^{10} c_0, s4=10^5 c_0,s_2=
10 c_0 $ in Fig. 9; $s_7=10^{10} c_0$ in Fig. 10; $s_7=10^{10} c_0,
s5=10^6 c_0$ in Fig. 11; $s_8=10^{12} c_0$ in Fig. 12; $s_8=10^{12}
c_0, s_6=10^5 c_0$ in Fig. 13, where $c_0=5+5\mathbf{i}$. In
particular, Figs. 9 and 11 provide the first two non-trivial
examples of a complete decomposition associated with four levels.

\noindent{\bf Conclusion}  The central theme of this paper is an
attempt to elucidate how normal waves can evolve into a rogue wave.
It is well known that when a classical envelope soliton interacts
with a background plane wave, then a breather is formed
\cite{Kharif}. Thus, there exist different types of breathers,
depending on the various combinations of envelope solitons and
background plane waves. It has been predicted that the maximum wave
field generated due to the interaction of an envelope soliton with a
background plane wave, depends on the linear superposition between
the amplitudes of the soliton and the background plane wave. The
problem of early detection of  rogue waves is a challenging task.
Indeed, since the NLS breathers are homoclinic orbits, even the
slightest perturbation resulting from roundoff errors during
numerical simulation, can trigger a false rogue like behavior.
Akhmediev {\sl et al.} \cite{akhmediev11} have devised a model for
early detection of rogue waves in a chaotic field, which would help
marine travel in stormy conditions, as it would provide an early
warning system for  rogue waves. Just before the appearance of the
high-peak wave in real space, the spectra of unit patches of the
chaotic wave fields show a specific triangular feature. Thus, the
analysis of the formation of such specific features could help the
early detection of rogue waves.

The two conjectures described in this article elucidate the
formation of higher-order rogue waves. By understanding the
generating mechanism for higher order rogue waves as a result of the
fission and the fusion of $n$ degenerate breathers, the formation of
the desired  triangular pattern (and of a new class of circular
pattern reported in this paper) is a basic features of rogue waves,
which may have an important impact on their early detection. The
constructions of specific triangular and circular patterns provide
simple implementations of the generic results presented in this
paper.

 {\bf Acknowledgements} This work is
supported by the NSF of China under Grants No.10971109 and
No.11271210, the K.C.Wong Magna Fund in Ningbo University and the
Natural Science Foundation of Ningbo under Grant No.2011A610179. We
thank Professor Yishen Li(USTC,Hefei,China) for his useful
suggestions on rogue waves and also thank Shuwei Xu, Linling Li,
Lijuan Guo, Yongshuai Zhang for their help with figures. KP wishes
to thank the DST, DAE-BRNS, UGC and CSIR, Government of India, for
the financial support. A. S.F. is grateful to EPSRC,UK, and to the
Onassis foundation, USA, for their generous support.

\mbox{}
\begin{figure}[htb]
\begin{center}
\includegraphics*[height=8 cm, width=8 cm]{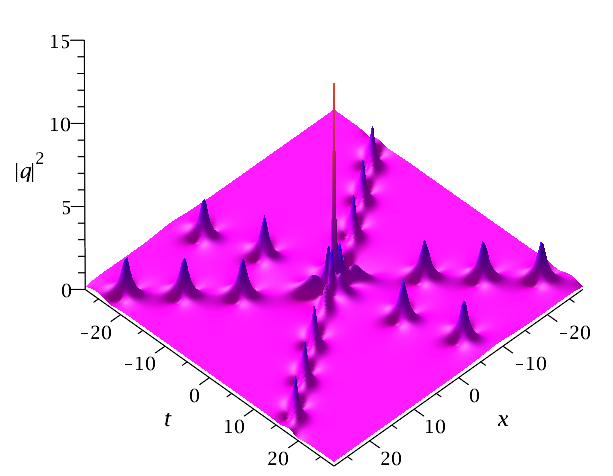}\mbox{}\\ 
\includegraphics*[height=8 cm, width=8 cm]{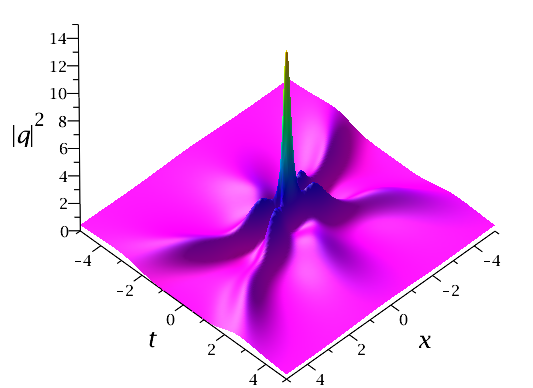} 
 \caption{{(Color online) The fusion and
fission of three synchronous breathers on the $(x,t)$ plane.  The
lower panel is a local profile in the interaction area of the upper
panel}}
\end{center}
\end{figure}
\begin{figure}[htb]
\begin{center}
\includegraphics*[height=8 cm, width=8 cm]{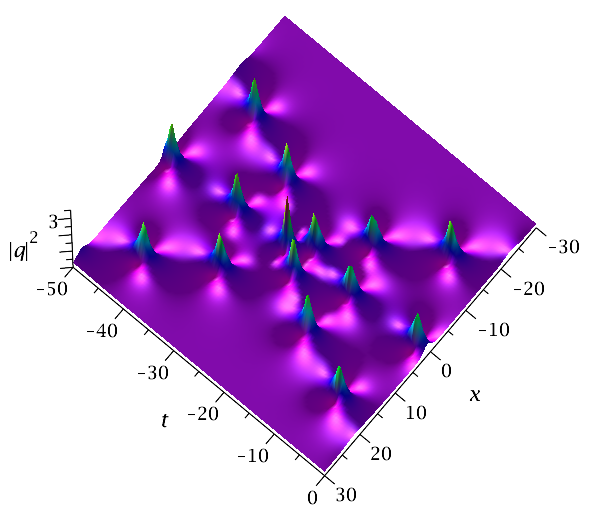}\\ 
\includegraphics*[height=8 cm, width=8 cm]{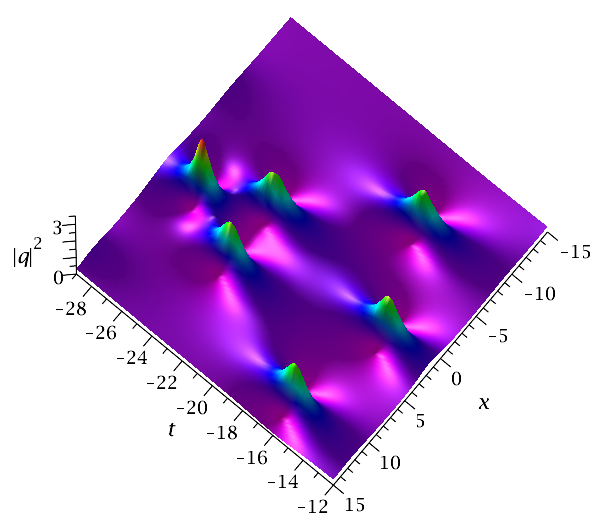} 
 \caption{{(Color online) The fusion and
fission of three anti-synchronous breathers on the $(x,t)$ plane.
The lower panel is a local triangle pattern in the interaction area
of the upper panel.}}
\end{center}
\end{figure}
\begin{figure}[htb]
\begin{center}
\includegraphics*[height=8 cm, width=8 cm]{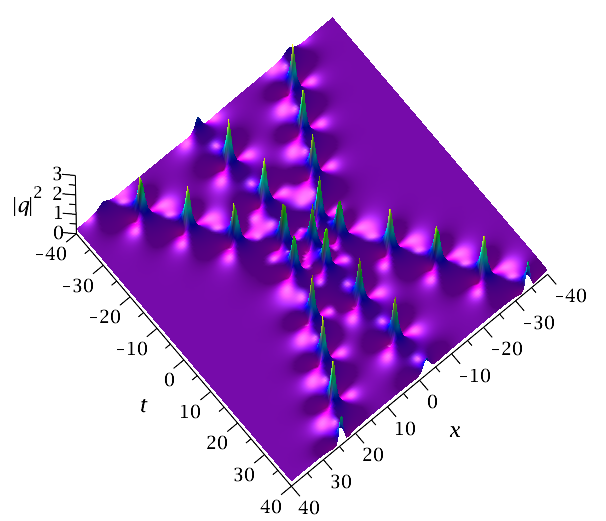}\\  
\includegraphics*[height=8 cm, width=8 cm]{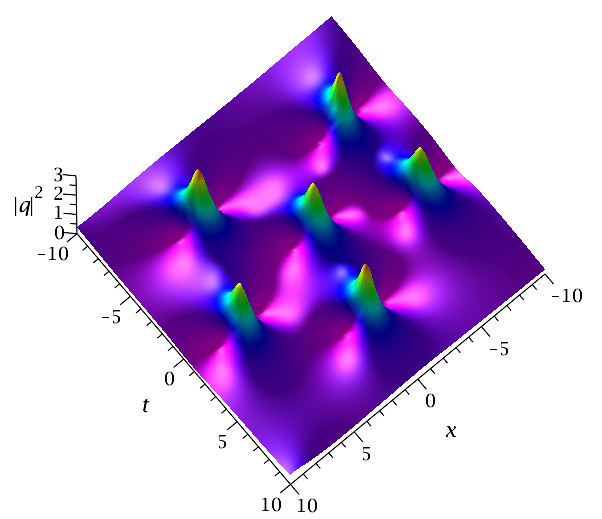} 
 \caption{{(Color online) The fusion and
fission of three anti-synchronous breathers on the $(x,t)$ plane.
The lower panel is a local circular pattern in the interaction area
of the upper panel.}}
\end{center}\mbox{}\vspace{-1cm}
\end{figure}
\begin{figure}[htb]
\begin{center}
\includegraphics*[height=8 cm, width=8cm]{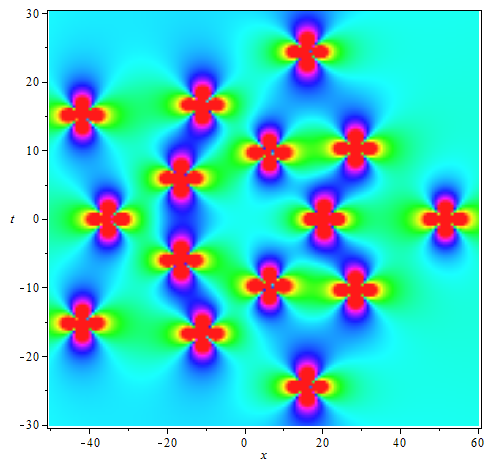} \\ 
\includegraphics*[height=8 cm, width=8cm]{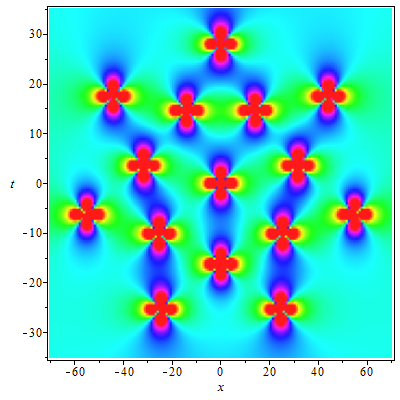} 
 \caption{(Color online) The polygon pattern
of an order-5 RW.  
The upper pentagon has three concentric circles, and each of them has
five peaks. The lower heptagon has two  concentric circles,
and each of them has seven peaks.}
\end{center}
\end{figure}
\begin{figure}[htb]
\begin{center}
\includegraphics*[height=8 cm, width=8cm]{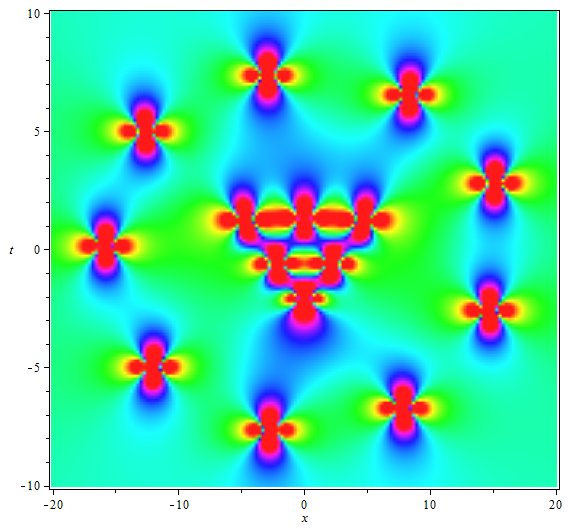}\\ 
\includegraphics*[height=8 cm, width=8 cm]{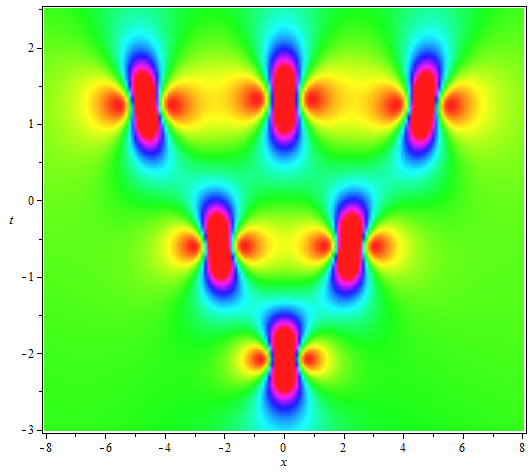} 
 \caption{(Color online) A triangle pattern in
a circle for an order-5 RW. The lower panel is a local central
profile of the upper panel.}
\end{center}
\end{figure}
\begin{figure}[htb]
\begin{center}
\includegraphics*[height=8 cm, width=8 cm]{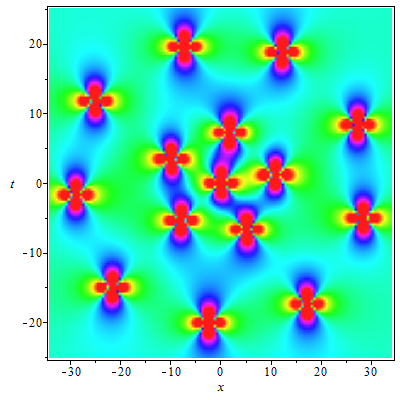}\\ 
\includegraphics*[height=8 cm, width=8 cm]{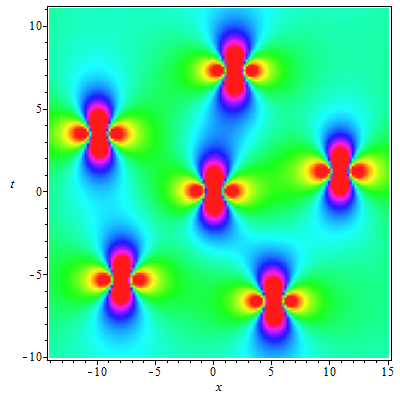} 
 \caption{(Color online) Decomposition of an
order-5 RW. The lower panel is a local central profile of the upper
panel.}
\end{center}\mbox{}\vspace{-2cm}
\end{figure}
\begin{figure}[ht]
\begin{center}
\includegraphics*[height=8 cm, width=8 cm]{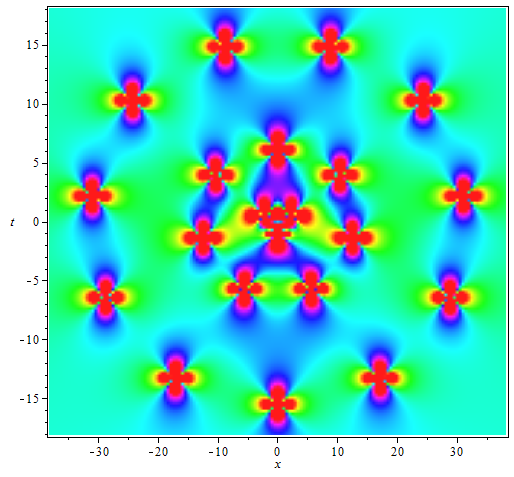}\\ 
\includegraphics*[height=8 cm, width=8 cm]{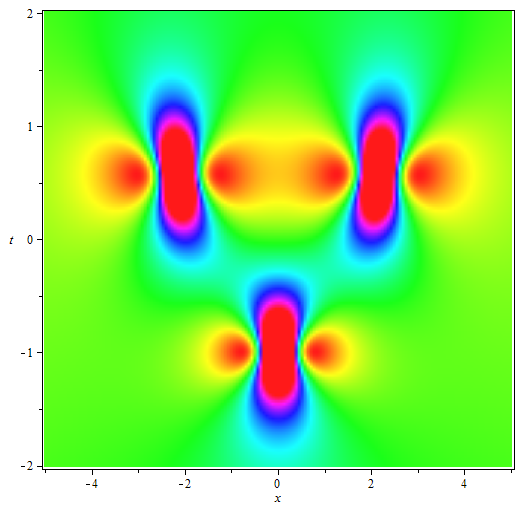} 
 \caption{(Color online) Decomposition of an
order-6 RW. The lower panel is a local central profile of the upper
panel.}
\end{center}
\end{figure}

\begin{figure}[ht]
\begin{center}
\includegraphics*[height=8 cm, width=8 cm]{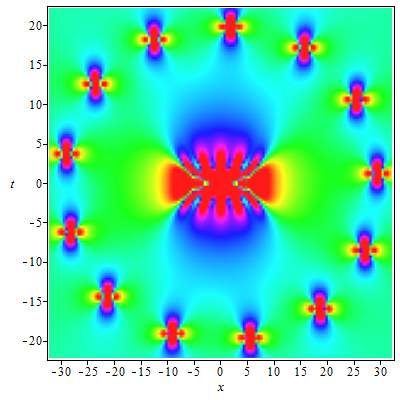}\\ 
\includegraphics*[height=8 cm, width=8 cm]{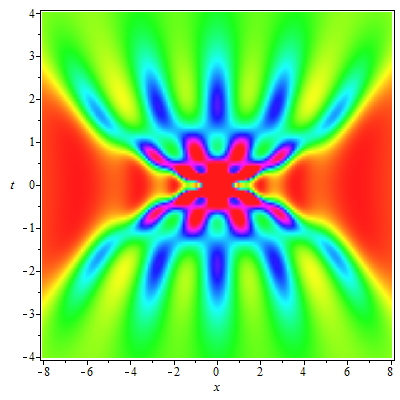} 
 \caption{(Color online) Decomposition of an
order-7 RW. The lower panel is a local central profile of the upper
panel.}
\end{center}
\end{figure}

\begin{figure}[ht]
\begin{center}
\includegraphics*[height=8 cm, width=8 cm]{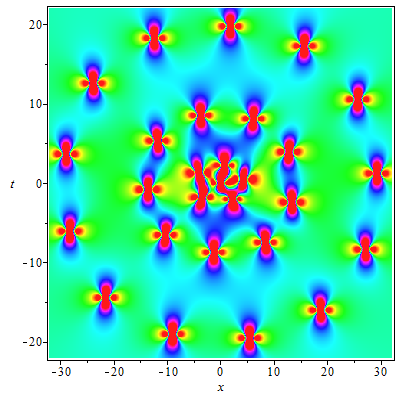}\\ 
\includegraphics*[height=8 cm, width=8 cm]{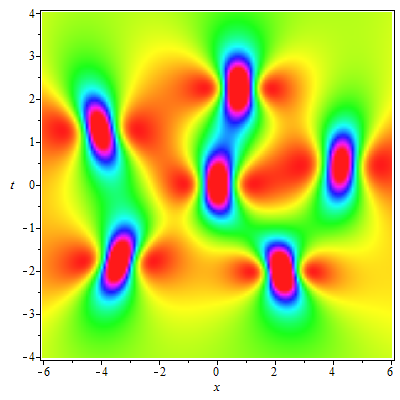} 
 \caption{(Color online) Decomposition of an
order-7 RW. The lower panel is a local central profile of the upper
panel.}
\end{center}
\end{figure}

\begin{figure}[ht]
\begin{center}
\includegraphics*[height=8 cm, width=8 cm]{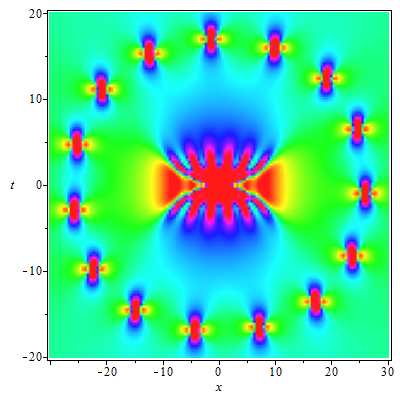}\\ 
\includegraphics*[height=8 cm, width=8 cm]{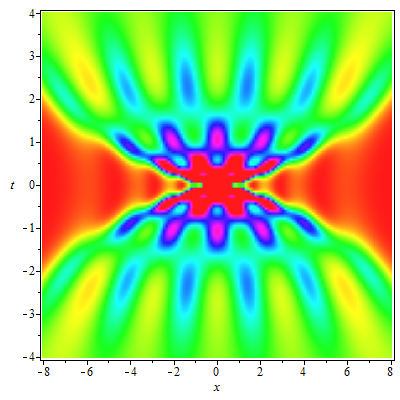} 
 \caption{(Color online) Decomposition of an
order-8 RW. The lower panel is a local central profile of the upper
panel.}
\end{center}
\end{figure}

\begin{figure}[ht]
\begin{center}
\includegraphics*[height=8 cm, width=8 cm]{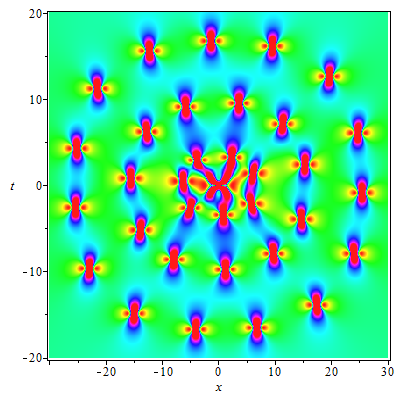}\\ 
\includegraphics*[height=8 cm, width=8 cm]{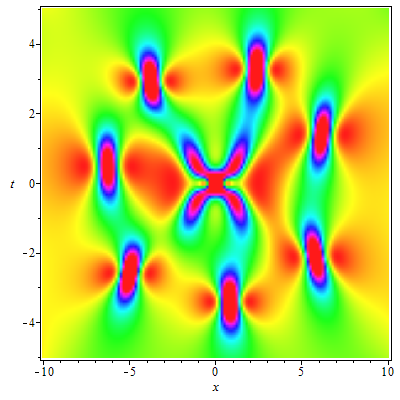} 
 \caption{(Color online) Decomposition of an
order-8 RW. The lower panel is a local central profile of the upper
panel.}
\end{center}
\end{figure}

\begin{figure}[ht]
\begin{center}
\includegraphics*[height=8 cm, width=8 cm]{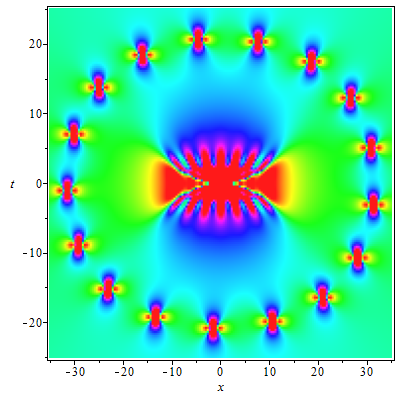}\\
\includegraphics*[height=8 cm, width=8 cm]{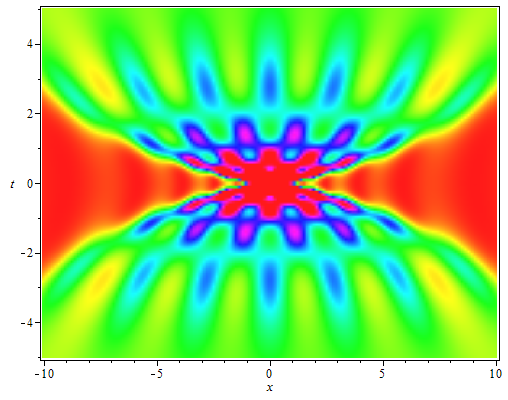}
 \caption{(Color online) Decomposition of an
order-9 RW. The lower panel is a local central profile of the upper
panel.}
\end{center}
\end{figure}
\begin{figure}[ht]
\begin{center}
\includegraphics*[height=8 cm, width=8 cm]{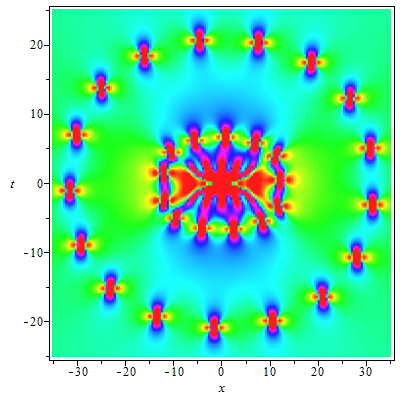}\\
\includegraphics*[height=8 cm, width=8 cm]{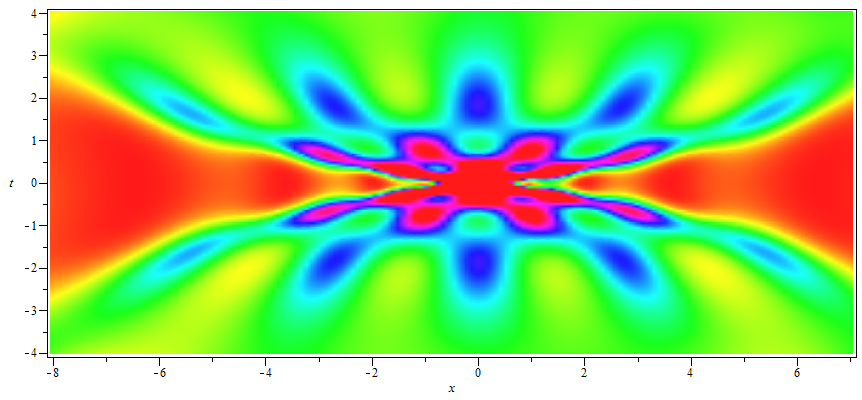}
 \caption{(Color online) Decomposition of an
order-9 RW. The lower panel is a local central profile of the upper
panel.}
\end{center}
\end{figure}

\end{document}